\documentclass[structabstract]{aa}  
\usepackage{graphicx}
\usepackage{txfonts}
\begin{document}
\title{Puzzling asteroid 21 Lutetia: our knowledge prior to the Rosetta fly-by
\thanks{Based on observations carried out at the ESO-NTT (La Silla, Chile), the Telescopio Nazionale Galileo
(La Palma, Spain), the Crimean Astrophysical Observatory (Ukraine), the Asiago Astrophysical Observatory (Italy) and Complejo Astron\'{o}mico El Leoncito (Casleo, Argentine).}}

%\subtitle{}

   \author{I.N. Belskaya \inst{1,2},
          S. Fornasier \inst{1,3},
          Yu.N. Krugly \inst{2},
          V.G. Shevchenko \inst{2},
          N.M. Gaftonyuk \inst{4},
          M.A. Barucci \inst{1},
          M. Fulchignoni \inst{1,3},
          R. Gil-Hutton \inst{5}}

\offprints{I.N. Belskaya}

\institute{LESIA, Observatoire de Paris, France\\
        \email{irina.belskaya@obspm.fr}
\and Institute of Astronomy of Kharkiv Karazin National University, Kharkiv,
     Ukraine
\and Universit\'{e} de Paris 7 Denis Diderot, France
\and Crimean Astrophysical Observatory, Crimea, Simeiz, Ukraine
\and Complejo Astron\'{o}mico El Leoncito (Casleo) and Universidad Nacional de San Juan, San Juan, Argentina
}

   \date{Received 5 January, 2010 / Accepted 2 March, 2010}

% \abstract{}{}{}{}{}
% 5 {} token are mandatory

  \abstract
   {}
   {A wide observational campaign was carried out in 2004-2009 aimed to complete the
   ground-based investigation of Lutetia prior to the Rosetta fly-by in July 2010.}
   {We have obtained BVRI photometric and V-band polarimetric measurements over a wide range of phase angles, and visible and infrared spectra in the 0.4-2.4 $\mu$m range. We analyzed them together with previously published data to retrieve information on Lutetia's surface properties.}
   {Values of lightcurve amplitudes, absolute magnitude, opposition effect, phase coefficient and BVRI colors of Lutetia surface seen at near pole-on aspect have been determined. We defined more precisely parameters of polarization phase curve and showed their distinct deviation from any other moderate-albedo asteroid. An indication of possible variations both in polarization and spectral data across the asteroid surface was found. To explain features found by different techniques we propose that (i) Lutetia has a non-convex shape, probably due to the presence of a large crater, and heterogeneous surface properties probably related to surface morphology; (ii) at least part of the surface is covered by a fine-grained regolith with particle size less than 20 $\mu$m; (iii) the closest meteorite analogues of Lutetia's surface composition are particular types of carbonaceous chondrites or Lutetia has specific surface composition not representative among studied meteorites.}
   {}


\titlerunning {21 Lutetia: our knowledge prior to the Rosetta fly-by}
\authorrunning{I.N. Belskaya et al.}

   \maketitle
%
%________________________________________________________________

\section{Introduction}

Asteroid 21 Lutetia has been extensively observed by different techniques for more than 30 years. Initially the interest in this object was connected with its classification as an M-type asteroid with a possible metallic composition (see Bowell et al. 1978). Since 2004 when Lutetia was selected as a target of the Rosetta mission, the volume of observational data on this asteroid is rapidly growing (see Barucci and Fulchignoni 2009 for a review). 

On the basis of photometric data obtained in 1962-1998 Torppa et al. (2003) determined the pole coordinates $\lambda$$_{p}$ =39$^o$ (220$^o$), $\beta$$_{p}$=3$^o$ and the sidereal rotation period P$_{sid}$=8.165455 h. The shape was found to have some irregular features with rough global dimensions a/b = 1.2 and b/c = 1.2.  Recently Drummond et al. (2009) gave new estimations of these parameters including in analysis adaptive optics images of Lutetia at the Keck telescope: $\lambda$ $_{p}$=49$^o$, $\beta$$_{p}$=-8$^o$ and a shape of 132x101x76 km with formal uncertainties of 1 km for the equatorial dimensions, and 31 km for the shortest axis. 

On the basis of spectral and polarimetric observations, three types of meteorites are generally taken into consideration as possible analogues: iron meteorites (Bowell et al. 1978, Dollfus et al. 1979), enstatite chondrites (Chapman et al. 1975, Vernazza et al. 2009) and some types of carbonaceous chondrites, mainly CO3 or CV3 (Belskaya \& Lagerkvist 1996, Birlan et al. 2004, Barucci et al. 2008, Lazzarin et al. 2009). The main problem in spectral data interpretation is the featureless spectrum of Lutetia. A presence of few minor features in the visible range was reported and interpreted as indicative of aqueous alteration material consistent with carbonaceous chondrites composition (see Lazzarin et al. 2009 and references therein). A 3 $\mu$m feature associated with hydrated minerals was found by Rivkin et al. (2000). In the emissivity spectra a narrow 10 $\mu$m emission feature was found (Feierberg et al. 1983, Barucci et al. 2008). It was interpreted as indicative the presence of fine silicate dust (Feierberg et al. 1983). According to Barucci et al. (2008) the emissivity spectrum is similar to that of the CO3 and CV3 carbonaceous chondrites with a grain size less than 20 $\mu$m.

To constrain surface composition knowledge of Lutetia's albedo is considered to be very important. However up to now the diversity in albedo estimations by different techniques is quite large, spanning from 0.1 (Zellner et al. 1976) to 0.22 (Tedesco et al. 2002).  Polarimetric method of albedo determination gives contradictory results (Zellner et al. 1976, Gil-Hutton 2007). An accuracy of radiometric albedo strongly depends on adopted absolute magnitude which is not well-determined for Lutetia due to lack of observations at small phase angles.

In this paper we present new photometric, polarimetric and spectral observations of Lutetia carried out in 2004-2009. They were aimed to determine absolute magnitude and albedo, and to put additional constrains on surface properties. Their analysis together with previously published data is given. We think that such analysis is important not only for deriving physical characteristics of this particular asteroid but first of all for checking the efficiency of remote techniques in the study of atmosphereless bodies. 

\section{Observations and results}
\subsection{Photometry}
The observations were carried out in 2004 using the 0.7-m telescope of Chuguev Observational Station situated 70 km from Kharkiv, and in 2008-2009 using the 1-m telescope of the Crimean Astrophysical Observatory in Simeiz, Crimea. The 0.7-m telescope was equipped with a SBIG ST-6 UV camera mounted in the Newtonian focus (f/4). In Simeiz we used a SBIG ST-6 camera placed on the 1-m Ritchey-Chretien telescope equipped with a focal reducer (f/5 system). The photometric reduction of the CCD frames was performed by using the ASTPHOT package developed at DLR by S. Mottola (Mottola et al. 1994). The absolute calibration was done using standard stars with colors close to the solar ones from Landolt (1983, 1992) and Lasker et al. (1988). The measurements were made in the standard Johnson-Cousins photometric system. The method of CCD observations and data processing included all standard procedures was described in detail by Krugly et al. (2002). The mean time of observations in UT, the heliocentric (\textit{r}) and geocentric ($\Delta$) distances, the solar phase angle ($\alpha$), the ecliptic longitude ($\lambda$) and latitude ($\beta$) in epoch J2000.0, the magnitude V$_{0}$(1,$\alpha$) reduced to the lightcurve primary maximum and its estimated error, and finally photometric bands of observations are given in Table 1. The estimated error of the absolute photometry includes both uncertainty on photometric reduction, typically 0.01-0.02$^m$, and uncertainty on the lightcurve amplitude correction. 
\begin{table}[ht]
\caption{Aspect data of photometric observations and magnitudes}    % title of Table
\label{table:1}      % is used to refer this table in the text
       \scriptsize{
\centering
\begin{tabular}{ccccrccc}        % centered columns (4 columns)
\hline\hline                 % inserts double horizontal lines
Date &   r   & $\Delta$ & $\lambda$ & $\beta$ & $\alpha$ & V$_{0}$(1,$\alpha$) & Filter\\    % table heading
(UT)     & (AU)  &   (AU)   &   (deg)  & (deg)  & (deg)  & (mag) &  \\
\hline                        % inserts single horizontal line
2004 09 16.07&2.163&1.416&47.72&-3.86&22.17&8.32$\pm$0.03&BVRI \\
2004 09 17.06&2.164&1.408&47.74&-3.86&21.87&8.35$\pm$0.04&BVRI \\
2004 10 07.06&2.195&1.285&46.31&-3.93&14.12&8.06$\pm$0.03&BVRI \\
2004 10 08.09&2.197&1.281&46.15&-3.93&13.65&8.04$\pm$0.02&BVRI \\
2004 11 10.81&2.254&1.274&38.40&-3.34& 4.75&7.63$\pm$0.02&V  \\
2008 11 28.96&2.420&1.434&69.00&-1.15& 0.91&-            &V  \\
2008 11 29.76&2.421&1.435&68.79&-1.12& 0.58&7.28$\pm$0.02&V  \\
2008 11 30.91&2.423&1.437&68.50&-1.10& 0.51&7.24$\pm$0.02&V  \\
2008 12 01.96&2.425&1.440&68.22&-1.08& 0.89&7.30$\pm$0.02&V  \\
2008 12 02.99&2.427&1.442&67.95&-1.05& 1.39&7.36$\pm$0.02&V  \\
2008 12 03.86&2.429&1.445&67.74&-1.03& 1.81&7.40$\pm$0.02&V  \\
2008 12 15.71&2.450&1.499&64.80&-0.74& 7.59&7.78$\pm$0.02&V  \\
2009 03 10.80&2.593&2.569&70.60& 0.63&22.18&8.28$\pm$0.03&VR  \\
2009 03 11.75&2.594&2.583&70.87& 0.64&22.12&8.25$\pm$0.03&VR  \\
\hline
\end{tabular}
}
\end{table}

During our observations in 2004 we were not able to cover small phase angles due to bad weather conditions and the observational program was continued in 2008. According to the recent estimates of Lutetia's pole coordinates $\lambda$$_{p}$ =51$^o$ (220$^o$), $\beta$$_{p}$=-4$^o$(B. Carry, personal communication) all our observations were made close to the pole-on direction with an aspect angle $\approx$10$^o$ in 2004 and $\approx$20$^o$ in 2008-2009. The composite lightcurves for each apparations are shown in Fig.1 and 2. The lightcurve amplitude increased from 0.06$^m$ in 2004 to 0.09$^m$ in 2008 and 0.12$^m$ in 2009 at the phase angle as large as 22$^o$. The lightcurves show an irregular behaviour with one pair of extrema. The measured lightcurve amplitudes and lightcurve features are consistent with the observations of 1981 (Lupishko et al. 1983, Zappala et al. 1984) and 1985 (Lupishko et al. 1987, Dotto et al. 1992), obtained near pole-on aspect too. 

\begin{figure}[t]
      \centering
   \includegraphics[angle=0,width=7.7cm]{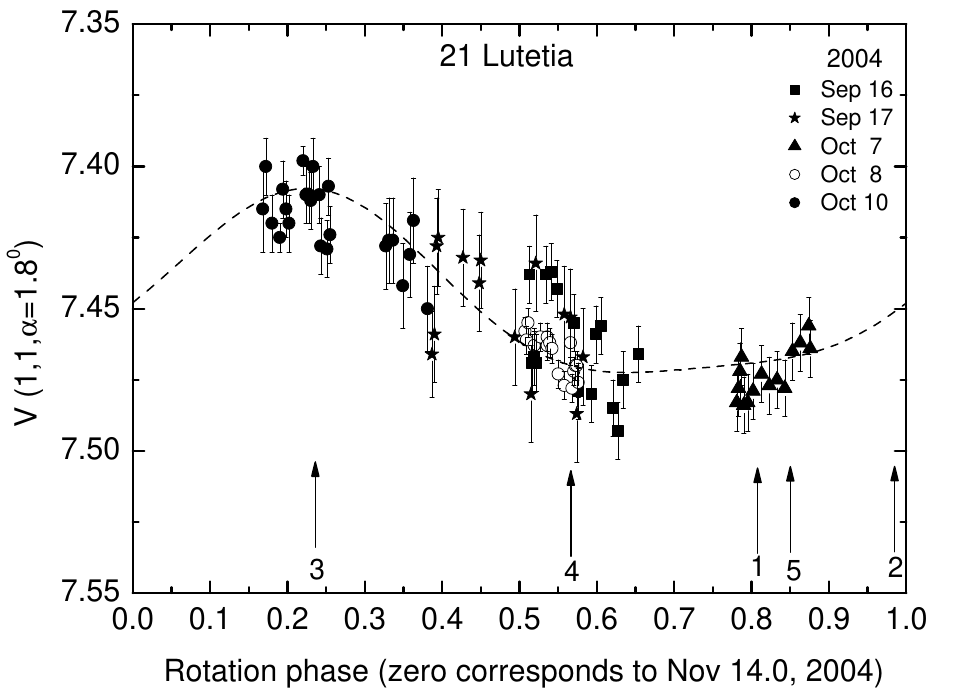}
\caption{Composite lightcurves of 21 Lutetia in 2004 apparition fitted with the Fourier fit. The arrows indicate rotation phases of our spectral observations (see Table 3 and Fig.7).}
%             \label{snl}%
\end{figure}
%\begin{figure}
\begin{figure}[t]
      \centering
   \includegraphics[angle=0,width=7.7cm]{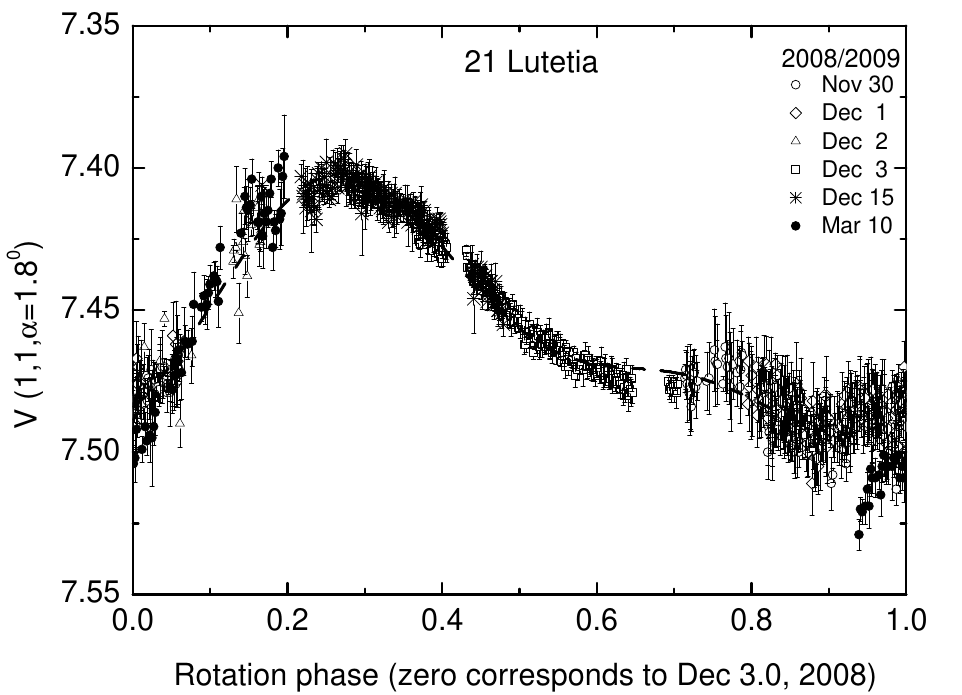}
\caption{Composite lightcurves of 21 Lutetia in 2008/2009 apparition fitted with the Fourier fit.}
%             \label{snl}%
\end{figure}
    
To obtain the phase function we normalized all the data to the lightcurve primary maximum. Errors due to amplitude corrections were taken into account in the magnitude's uncertainties. We also used the V-magnitudes measured in 2004 at the phase angle of 27.4$^o$ by Mueller et al. (2006) and normalized it to the lightcurve maximum using our lightcurve for the 2004 opposition. We applied the same procedure to the available observations of Lutetia from the 1981, 1983, and 1985 oppositions, separately for each opposition. These data were obtained by different authors (Lupishko et al. 1983, 1987, Dotto et al. 1992, Lagerkvist et al. 1995, Zappala et al. 1984) at a variety of phase angles and were not analyzed jointly. For analysis we used an updated value of Lutetia's sidereal rotation period P$_{sid}$=8.168268 h and normalized all the data to the same maximum. 

It was found that observations in the four oppositions corresponding to pole-on aspect are mutually in good agreement within the error bars. The obtained phase function is shown in Fig.3. Fitting the data with the HG fit (Bowell et al. 1989) and with the linear-exponential fit (Kaasalainen et al. 2003) we obtain practically identical curves. The HG-fit to the phase curve normalized to the lightcurve primary maximum gives H=7.20$\pm$0.01 and G=0.12$\pm$0.01. Note, that for the phase function normalized to the mean lightcurve H=7.25$\pm$0.01. The phase coefficient obtained by the linear fit to the data at phase angles $\alpha$$\geq$7$^o$ is equal to $\beta$=0.034$\pm$0.001$^m$/deg and the magnitude at zero phase angle corresponding to the extrapolation of the linear fit is V(1,0)=7.56$\pm$0.01. The amplitude of the opposition effect defined as an increase in magnitude above the linear fit at zero phase angle was estimated to be 0.36$^m$. Both the opposition effect amplitude and the value of the phase slope are consistent with moderate-albedo surface. Based on the empirical correlation between phase coefficient and albedo (Belskaya and Shevchenko 2000) an average albedo in the range of 0.12-0.20 is expected for Lutetia's surface. 

\begin{figure}[t]
     \centering
   \includegraphics[angle=0,width=9cm]{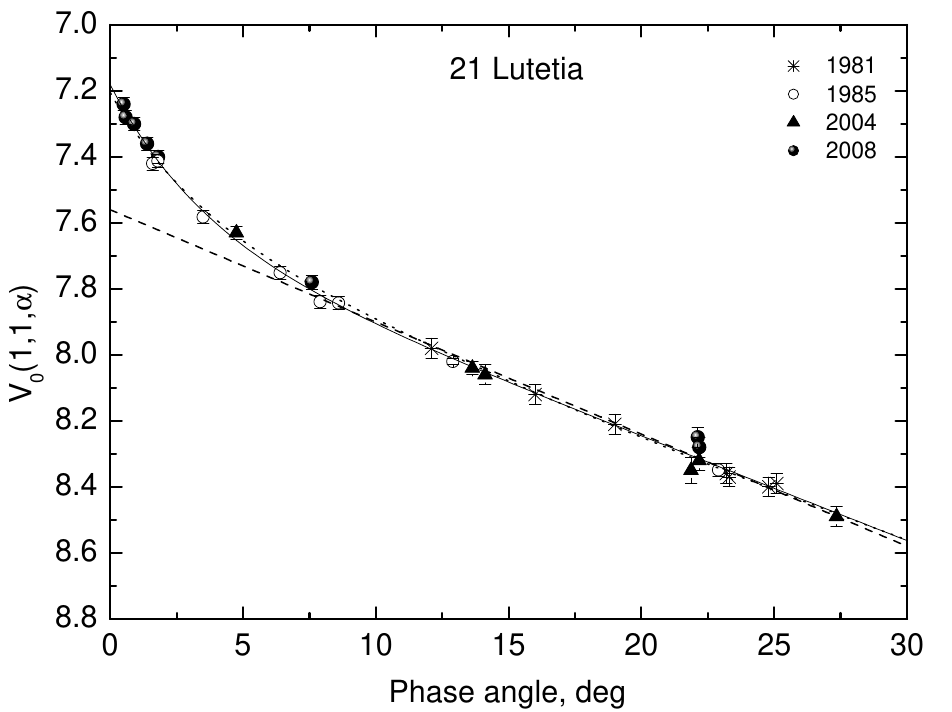}
\caption{Magnitude phase dependence for Lutetia based on observations at different apparitions at near polar aspects fitted by the HG function (the dotted line) and the linear-exponential function (the solid line). The dashed line shows liner fit to the data at $\alpha$$\geq$7$^o$. }
%             \label{snl}%
\end{figure}
 
\begin{figure}[t]
      \centering
   \includegraphics[angle=0,width=9cm]{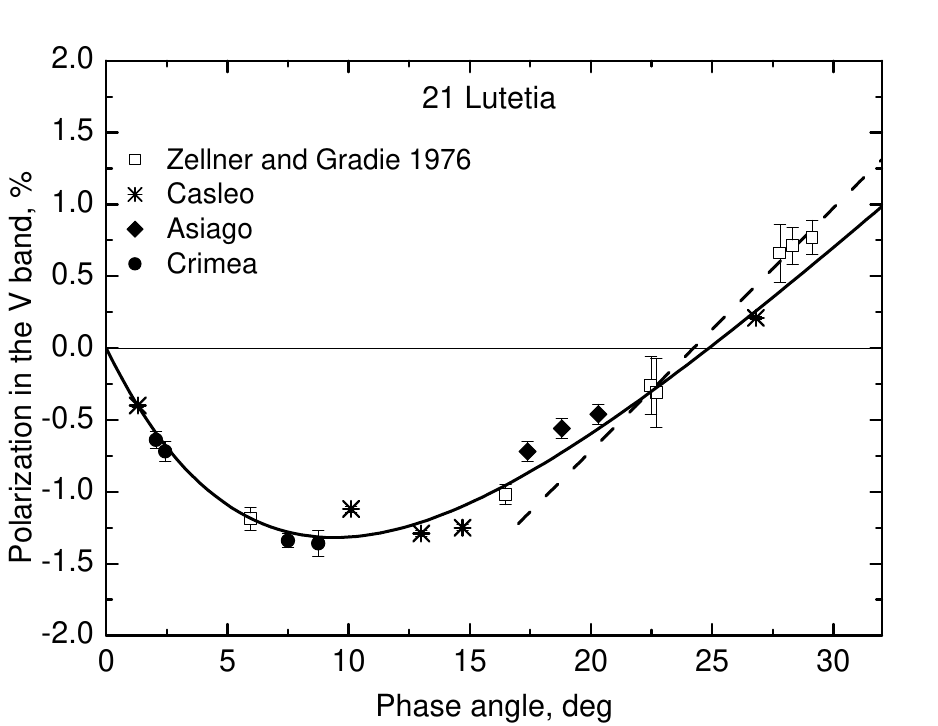}
\caption{Polarization phase dependence for Lutetia based on observations in 1973-2008 at different observational sites fitted by the linear-exponential function (solid line). The dashed line indicates the polarimetric slope h as defined by Zellner and Gradie (1976).}
%              \label{snl}%
\end{figure}

The phase function obtained for the observations in 1983 at near-equatorial aspects is characterized by systematically lower magnitudes well fit by the HG-function with H=7.29$\pm$0.02 and G=0.13$\pm$0.03.  Thus, the difference between the absolute magnitudes at near polar and near equatorial aspects is found to be as small as 0.1$^m$. It implies that an upper limit of Lutetia's shape elongation b/c $\leq$ 1.1 in the case of homogeneous surface albedo. 

We also measured BVRI colors of Lutetia at different phase angles and found a slight increasing trend toward larger phase angles, not exceeding a level of 0.001$^m$/deg. The mean measured colors are B-V=0.65$\pm$0.01, V-R=0.42$\pm$0.01, and V-I=0.76$\pm$0.01.

\subsection{Polarimetry}

The first polarimetric observations of Lutetia were made in 1973 by Zellner and Gradie (1976). They derived a polarimetric slope \textit{h}=0.169\%/deg in the green filter with an effective wavelength of 0.52 $\mu$m and an albedo of 0.10 based on the empirical relationship "\textit{h}-albedo". They also measured the inversion angle  $\alpha$$_{inv}$=24.2$^o$ which appeared to be the largest one among all asteroids in their data-set. Polarimetric observations of Lutetia were successively carried out in 1985 in UBVRI filters at a phase angle of 7.5 deg close to the polarization minimum (Belskaya et al. 1987). They showed that the depth of polarimetric minimum reached 1.3\% in the V band slightly increasing with wavelength. Other observations of Lutetia were carried out in the framework of a coordinate program at three observatories: the Crimean Astrophysical Observatory (Ukraine), the Asiago Observatory (Italy) and Complejo Astron\'{o}mico El Leoncito (Casleo, Argentine), in order to cover phase angles which were not previously observed. Part of these data has been published among results of observations at each telescope (Fornasier et al. 2006, Gil-Hutton 2007, Belskaya et al. 2009). Here we report complementary observations of Lutetia not yet published. Table 2 presents the mean time of observations in UT, the phase angle $\alpha$, the polarization degree \textit{P} and position angle \textit{$\Theta$} in the equatorial coordinate system, together with the root-mean-square errors $\sigma$$_\textit{P}$ and $\sigma$$_\textit{$\Theta$}$, the calculated values of the corresponding \textit{P}$_{r}$ and position angle \textit{$\Theta$}$_{r}$ in the coordinate system referring to the scattering plane as defined by Zellner and Gradie (1976), and the telescope. Methods of observations and data processing were the same as described by Fornasier et al. (2006) for Asiago, Belskaya et al. (2009) for Crimea, and Gil-Hutton (2007) for Casleo.
\begin{table}[ht]
\caption{Results of polarimetric V-band observations of 21 Lutetia}    % title of Table
\label{table:2}      % is used to refer this table in the text
        \scriptsize{
\begin{tabular}{c r c c c c c c c}        % centered columns (4 columns)
\hline\hline                 % inserts double horizontal lines
Date &  $\alpha$   & P  & $\sigma$$_P$ & $\Theta$ & $\sigma$$_\textit{$\Theta$}$ & P$_{r}$ & $\Theta$$_{r}$ & Tel.\\    
% table heading
UT      & (deg)  &   (\%)   &   (\%)  & (deg)  & (deg)  & (\%) &  (deg) & \\
\hline                        % inserts single horizontal line
2004 10 17.96&	 8.75&	1.37&	0.09&	80.8&	1.9&	-1.36&	87.0&	1	\\
2006 04 06.79&  17.38&	0.81&	0.07&	97.0&	2.0&	-0.72&	78.1&	2 \\
2008 10 31.30&	14.80&	1.28&	0.02&	93.0&	0.5&	-1.25&	97.0&	3\\
2008 11 04.25&	13.10&	1.31&	0.02&	90.9&	0.4&	-1.29&	94.8&	3\\
\hline
\end{tabular}

1. 1.25 m, Crimea\\
2. 1.82 m, Asiago\\
3. 2.15 m, Casleo  
}
\end{table}
\begin{figure}[t]
      \centering
   \includegraphics[angle=0,width=7cm]{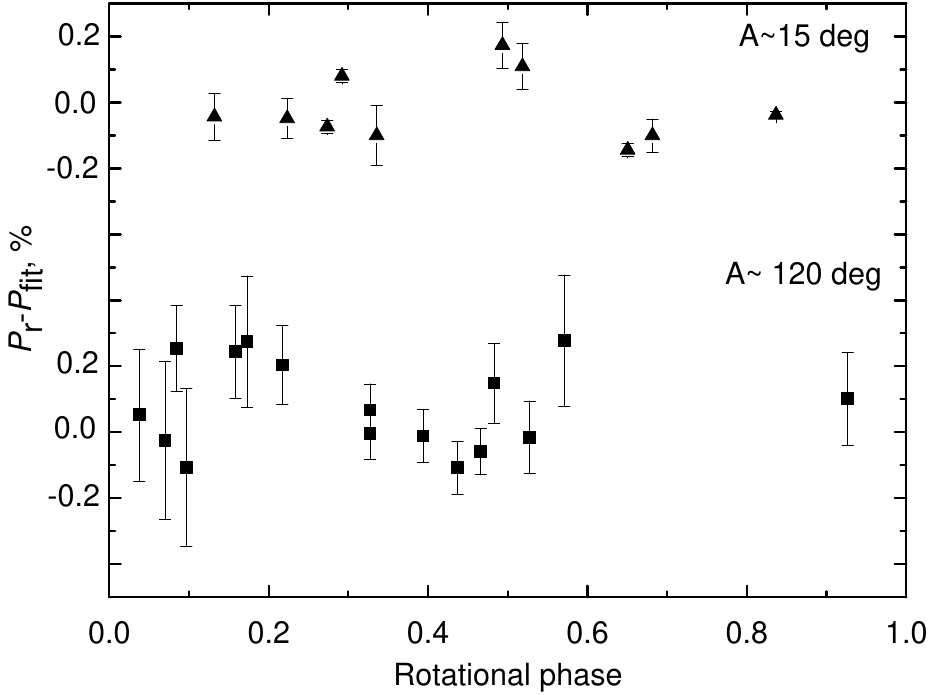}
\caption{The deviations of the polarization degree \textit(P$_r$) from the the linear-exponential fit to polarization phase curve \textit(P$_{fit}$) versus rotation phase at different aspect angle \textit{A}. }
%              \label{snl}%
\end{figure}
The polarization-phase function of Lutetia obtained using both new and published data is shown in Fig.4. The data were fit by the linear-exponential function as described by Kaasalainen et al. (2003). Similar curves are obtained fitting the data with the trigonometric fit (Lumme \& Muinonen 1993) or parabolic fit. The scatter of the data, rather large and exceeding the estimated errors of each measurement, may indicate a possible variation in polarization degree across the asteroid surface.  

We analyzed the deviations of the polarization degree from the fitted phase curve and found that they have a systematic rather than random nature. Fig.5 shows these deviations versus rotation phase for observations in 1973 corresponding to the aspect angle of about 120$^o$ and in 2004 and 2008 oppositions when the aspect was near pole-on (6-24$^o$). One can see that variations in the polarization degree tend to increase toward equatorial aspect and can reach up to 0.2\%.  

An amplitude of variations in polarization degree across the Lutetia's surface resembles that measured on asteroid 4 Vesta (e.g., Lupishko et al. 1988) and could be caused by the same reason, i.e. macroscale surface heterogeneity.  
The mean polarization phase dependence of Lutetia is characterized by the following parameters:
\textit{P}$_{min}$=-1.30$\pm$0.07\%, \textit{$\alpha$}$_{min}$=9.1$\pm$0.8 deg,  \textit{$\alpha$}$_{inv}$=25.0$\pm$0.4 deg, \textit{h}=0.131$\pm$0.009 \%/deg.

The polarimetric slope \textit{h} has a smaller value compared to that defined by Zellner \& Gradie (1976). It corresponds to the geometric albedo p$_V$= 0.13$\pm$0.02 when using the empirical relationship "\textit{h}-albedo", that was calibrated with albedos of meteorites (Zellner \& Gradie 1976), and  p$_V$=0.16$\pm$0.02 using the calibration based on IRAS albedos (Cellino et al. 1999). The difference between these two values is fully explained by different scales of albedos. Albedos of meteorites were measured in laboratory at phase angle $\alpha$ =5$^o$ while IRAS albedos were determined at zero phase angle using asteroid absolute magnitude H. 

\begin{figure}[t]
\centering
\includegraphics[angle=0,width=9cm]{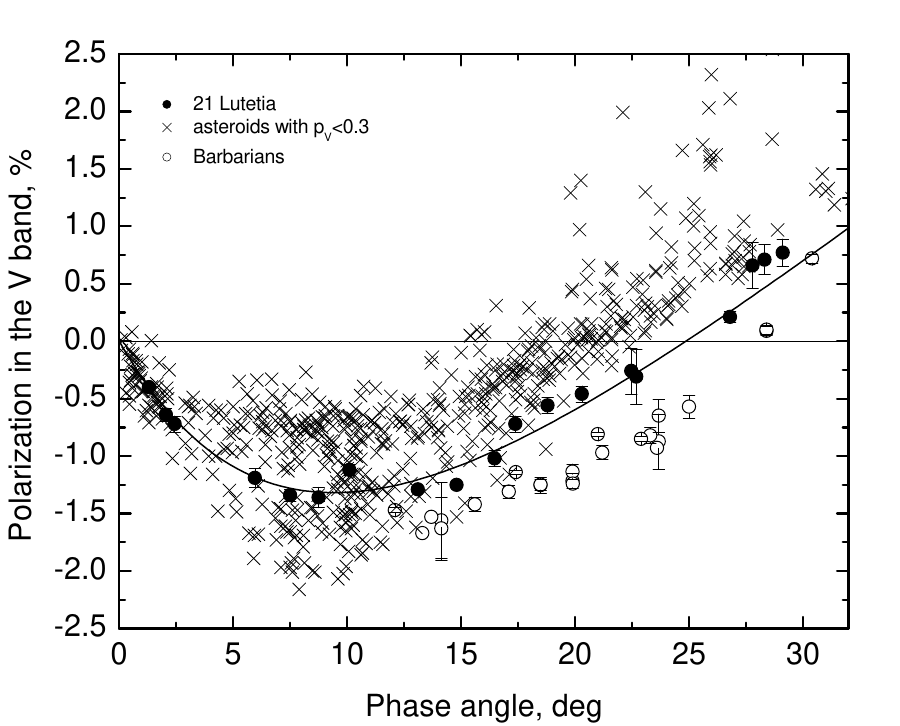}
\caption{The polarization-phase dependence of Lutetia (solid line) in comparison with available observations of moderate and low albedo asteroids (crosses) taken from the Asteroid Polarimetric Database (Lupishko \& Vasilyev 2008) and data for Barbara-like asteroids (circles) according with Cellino et al. (2007), Gil-Hutton et al. (2008), and Masiero \& Cellino (2009).}
\label{snl}%
\end{figure}

The most interesting polarimetric characteristic of Lutetia is its wide branch of negative polarization. Fig.6 shows Lutetia's data in comparison with available polarimetric measurements of low and moderate albedo asteroids. For comparison we used the Asteroid Polarimetric Database (Lupishko \& Vasilyev 2008) and we selected the data having an accuracy better than 0.2\% and concerning asteroids with albedo less than 0.3. Lutetia's observations characterized by the inversion angle as large as 25$^o$ represent a marginal case as compared to a variety of asteroids observed so far. Only asteroid 234 Barbara and four more asteroids called "Barbarians" show a polarization branch wider than that of Lutetia (see Fig.6). This group of moderate-albedo asteroids of the spectral types L, K or Ld exhibits anomalous polarization properties which have been interpreted as related to their specific surface composition (Cellino et al. 2006, Gil-Hutton et al. 2008, Masiero \& Cellino 2009). Note, that the value of polarization minimum of these asteroids considerably deviates from the well-known correlation "\textit{P}$_{min}$ - albedo" and can not be used for albedo estimation. In the case of Lutetia this correlation also fails. 

\begin{figure}[t]
\centering
\includegraphics[angle=0,width=9cm]{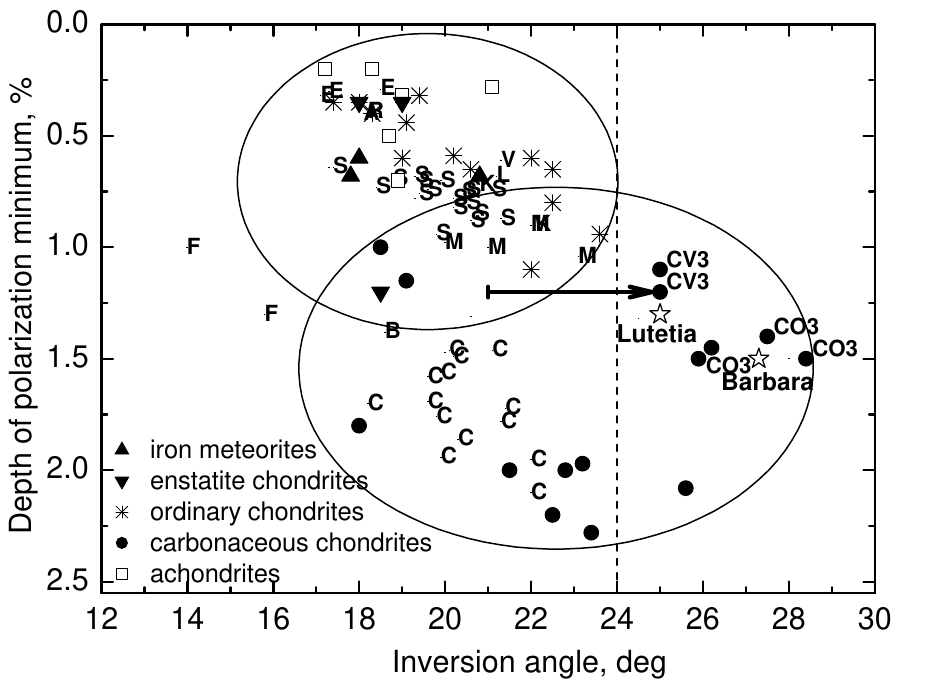}
\caption{Minimum polarization \textit{P}$_{min}$ (in absolute term) vs. inversion angle for asteroids and meteorites. The two ellipses outline the location of carbonaceous chondrites and all other types of meteorites. The arrow shows the changes in the inversion angle for the CV3 chondrite Allende when these angles are measured on a solid piece or on a crushed sample. Data for meteorites come from Zellner et al. (1977), Geake \& Dollfus (1986), Shkuratov et al. (1984), Lupishko \& Belskaya (1989). Asteroid polarimetric parameters were calculated fitting by linear-exponentional the data for individual asteroids contained in the Asteroid Polarimetric Database. Letters designate taxonomic class of asteroids according Tholen (1989).}
%\label{snl}%
\end{figure}

The presence of the negative polarization can be explained by several physical mechanisms, the most appropriate of which is considered to be the coherent backscattering mechanism (see Shkuratov et al., 1994 for a review) and the single particle scattering (e.g., Mu{\~n}oz et al. 2000). The coherent backscattering mechanism contributes both to the brightness opposition effect and to the negative polarization branch and is especially efficient for high albedo surfaces producing narrow backscattering peaks (Mishchenko et al. 2006). The measured phase curves of Lutetia (Fig.3 and 4) do not show any sharp features toward zero phase angle. Both phase curves are characterized by wide opposition effects which assume small relative contribution of the coherent backscattering. The contribution of the mechanism of the single particle scattering is still not well understood but its efficiency to produce wide negative polarization was shown by laboratory and numerical modeling (e.g. Mu{\~n}oz et al., 2000; Shkuratov et al. 2002). It was found that the negative branch becomes more prominent and the inversion angle increases in the case of a) increasing the refractive index, b) decreasing particle sizes down to sizes compared with wavelength, c) complex internal structure of particles, d) mixture of particles with high contrast in albedo (Mu{\~n}oz et al., 2000, Shkuratov et al. 1994, 2002, Zubko et al. 2005). One or several of the above-mentioned properties can be responsible for particular polarization characteristics of Lutetia. 

On the basis of the relationship between \textit{P}$_{min}$ and \textit{$\alpha$}$_{inv}$ Dollfus et al. (1975) mentioned that 21 Lutetia belonged to the group with a regolith of fines. This group was separated on the basis of the measurements of lunar fines having average grain sizes of the order of 10 $\mu$m with a range from less than 1 $\mu$m to several tens of microns (e.g., Geake \& Dollfus 1986). Later Lutetia's data were interpreted as indicative of a metallic surface with a grain size of 20-40 $\mu$m (Dollfus et al. 1979). The conclusion was based on measurements of specific powders, like titanium, dural, limonite, carbonyl iron globules, while neither pulverized iron meteorites nor pulverized enstatite chondrites match polarimetric curves of M-type asteroids. Laboratory measurements of iron meteorites and enstatite chondrites with particle sizes less than 50 $\mu$m show smaller inversion angles than that measured for Lutetia's surface (Lupishko and Belskaya 1989). A CV3 type of carbonaceous chondrites was mentioned as the best polarimetric analogue of Lutetia (Belskaya \& Lagerkvist 1996).  

Fig.7 shows an updated relationship between \textit{P}$_{min}$ and \textit{$\alpha$}$_{inv}$ for asteroids and meteorites.  Among meteorites the widest negative polarization branches are inherent for CV3 and CO3 types of carbonaceous chondrites. These types of chondrites are distinguished by relative abundances of refractory inclusions, in particular calcium-aluminum rich inclusions (CAI) (e.g. Scott \& Krot 2005). Sunshine et al. (2008) assumed that the presence in some CAIs of spinel, which has one of the highest indices of refraction among meteorite minerals, that may explain the large inversion angles. Another possible explanation is related to the fine structure of CV3 and CO3 meteorite samples measured with the polarimetric technique. The measurements of cleavage faces of solid pieces and pulverized samples for CV3 Allende and CO3 Kanzas chondrites (Shkuratov et al. 1984) showed that the depth of negative branch was practically the same for powder and solid samples while the inversion angle noticeably increased for a powder sample (see Fig. 7). However it is difficult to explain why pulverized samples of other types of carbonaceous chondrites show smaller inversion angles.  

At present laboratory measurements are available for rather limited sample of meteorites not covering all known meteorites classes. Any of measured iron meteorites, enstatite and ordinary chondrites did not show an inversion angle as large as found for Lutetia. Only particular types of carbonaceous chondrites are found to have wide negative polarization branch. It is possible that fine grained mixture of components with highly different optical properties (carbon, silicates, irons) is required to produce a large inversion angle seen for Lutetia.  

\subsection{Spectral observations}

The observations were made during two runs in November 2004 at the TNG telescope at la Palma, Spain and in January 2007 at the NTT telescope of the European Southern Observatory in Chile. 

At the TNG telescope we used the DOLORES spectrometer with two grisms:  the low resolution red grism (LR-R) covering the 0.51-0.95 $\mu$m range with a spectral dispersion of 2.9 A/px and the medium resolution blue grism MR-B, with a dispersion of 1.7 A/px covering the 0.4-0.7 $\mu$m range. The obtained spectra were separately reduced and then combined together to obtain the spectral coverage from 0.4 to 0.95 $\mu$m. For the infrared range we used the near infrared camera and spectrometer (NICS) equipped with an Amici prism disperser covering the 0.85-2.4 $\mu$m range. 

At the NTT telescope, visible spectra were acquired using the EMMI instrument with the grism covering the wavelength range of 0.41-0.96 $\mu$m with a dispersion of 3.1 A/px. The data acquisition and reduction techniques are described by Fornasier et al. (2008).  The observational circumstances are summarized in Table 3, which contains date and UT-time at the start of observations, the exposure time, telescope, instrument, airmass of the object, the name and airmass of solar analog star and the number corresponding to the rotation phase at the time of the observation, as shown in Fig.1. 

\begin{table}
%       \begin{center}
       \caption{Observational circumstances for spectral observations of 21 Lutetia}
        \label{table 3}
        \scriptsize{
\begin{tabular}{cccccccc}
 \hline
 Date &  UT-start  & T$_{exp}$ & Tel. & Instr. & Grism & Airm. &
Solar  \\
 & (hh:mm) & (s) & & &   & & analog \\
 \hline
2004 11 04 & 23:35 & 40 &  TNG & DOLORES & LR-R & 1.05 & 1 \\
2004 11 15 & 23:37 & 40 &  TNG & DOLORES & MR-B & 1.05 & 1 \\
2004 11 16 & 01:10 & 40 &  TNG & DOLORES & LR-R & 1.11 & 1 \\
2004 11 16 & 01:12 & 40 &  TNG & DOLORES & MR-B & 1.11 & 1 \\
2004 11 16 & 03:06 & 40 &  TNG & DOLORES & LR-R & 1.50 & 1 \\
2004 11 16  & 03:08 & 40 &  TNG & DOLORES & MR-B & 1.51 & 1 \\
2004 11 18  & 23:11 & 60 & TNG & NICS & AMICI   & 1.05  &1 \\
2007 01 20 & 08:47 & 120 & NTT & EMMI & GR1 & 1.55 & 2 \\
2007 01 20& 08:42 & 240 & NTT & EMMI & GR5 & 1.58 & 2 \\
\hline
\end{tabular}
%\end{center}
\\
1. Hyades64 (airmass 1.03)\\
2. La102-1081 (airmass 1.22) 
}
\end{table}
\begin{figure}[h]
      \centering
   \includegraphics[angle=0,width=7.7cm]{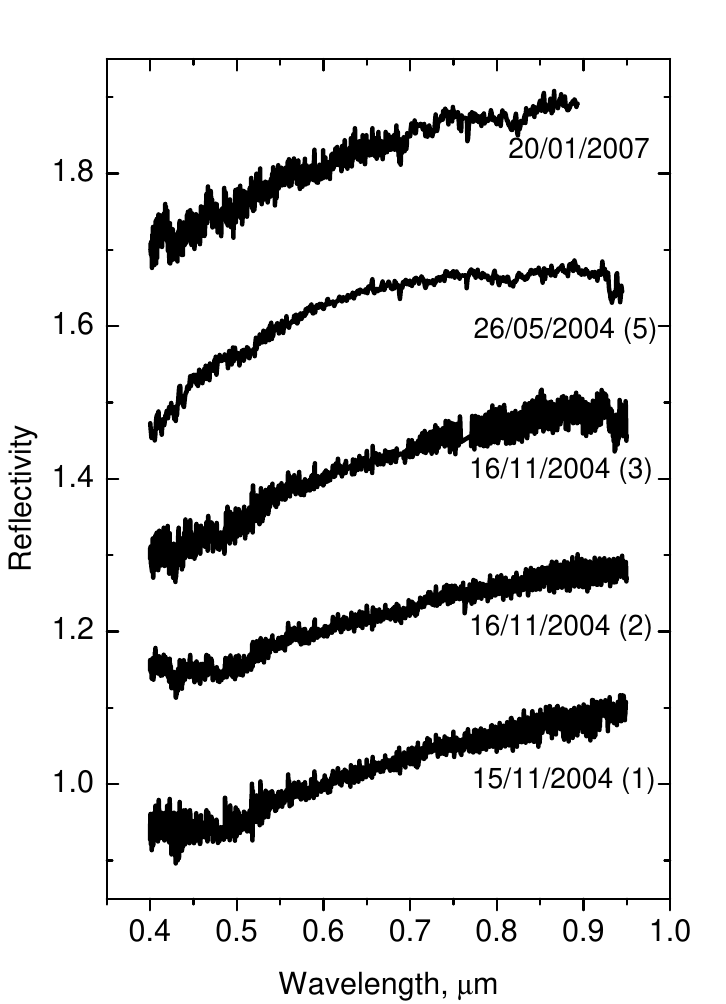}
              \label{snl}%
      \centering
   \includegraphics[angle=0,width=7.7cm]{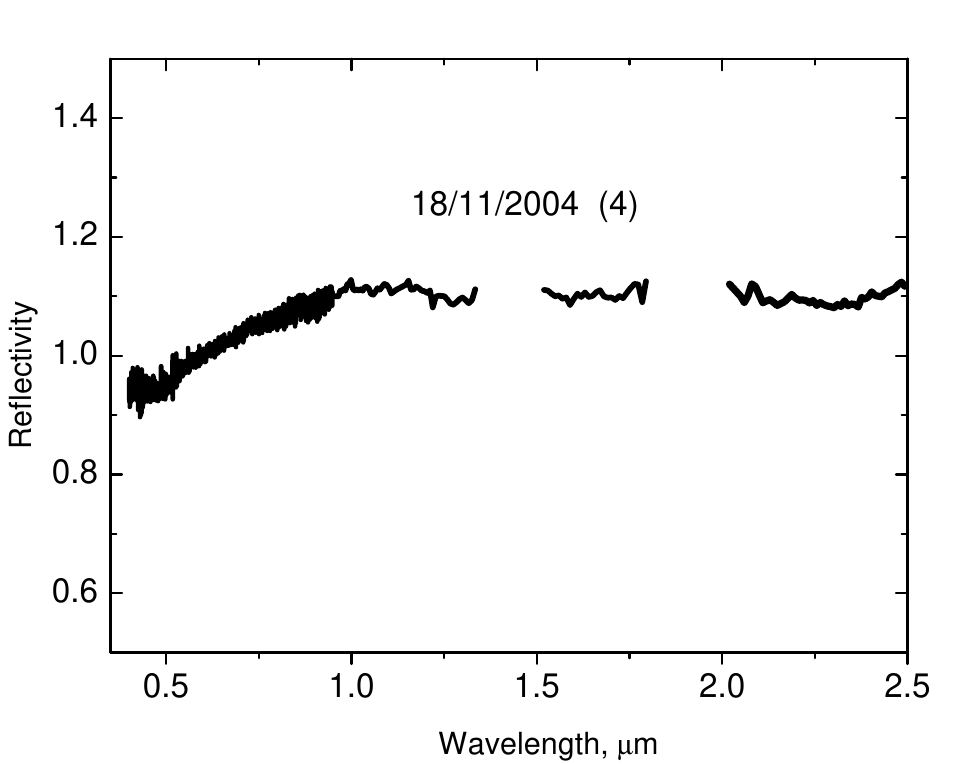}
\caption{Visual and near-infrared spectra of Lutetia.  The spectra were shifted by 0.2 for clarity. The numbers in the parentheses correspond to the spectrum numbers in Table 3 and Fig.1. The spectrum on 26/05/2004 is taken from Barucci et al. (2005).   }
              \label{snl}%
\end{figure}

The spectral data are presented in Fig. 8. Three visible spectra measured on Nov 15/16 ($\alpha$=7.3$^o$) at different rotation phase (see arrows in Fig.1) show noticeably different shape. In two spectra (1 and 2) a broad band at 0.45-0.55 $\mu$m is clearly visible while in the spectrum close to the lightcurve maximum (3) it becomes less evident. For comparison we also presented the spectrum taken on May 26, 2004 by Barucci et al. (2005) at near the same pole-on aspect but at larger phase angle ($\alpha$=24$^o$) which does not show a broad band at 0.45-0.55 $\mu$m. Such a band is also not seen in the spectrum taken in 2007. The spectrum corresponds to the opposite side of Lutetia with respect to the one covered by the spectra taken in 2004. The faint absorption around 0.83 $\mu$m seen in the spectrum is probably due to incomplete removal of telluric bands. On the other hand the faint absorption feature around 0.43 $\mu$m appears in all our spectra and seems to be real. 

The near-infrared spectrum measured on Nov 18, 2004 at the phase angle of $\alpha$=8.8$^o$ is flat with a small negative slope. It does not show any features detectable within the noise of the data. We compared it with the spectrum obtained by Birlan et al. (2006) at the same opposition but at larger phase angle $\alpha$=28.3$^o$ and did not find any phase angle effect. Both spectra have been taken quite at the same rotational phase are flat. The reddening with increasing
phase angle is not seen for Lutetia's surface in the spectral range of 0.8-2.5 $\mu$m.

Our visible spectra are in a good agreement with previous observations of Lutetia. We confirmed a presence of a broad feature at 0.45-0.55 $\mu$m in some of our spectra previously reported by Lazzarin et al. (2009) from observations in the same 2004 apparition. Lazzarin et al. (2009) suggested that the feature could be a superposition of several absorption bands caused by charge transfer involving various metal ions in pyroxenes. The faint absorption feature around 0.43 $\mu$m was not seen in the 2004 spectra by Lazzarin et al. (2009) but it was identified in some Lutetia's spectra obtained in 2000 (Busarev et al. 2004, Prokof'eva et al. 2005) and in 2003 (Lazzarin et al. 2004). The feature was interpreted as a possible indication of aqueous alteration activity (Lazzarin et al. 2004, 2009; Prokof'eva et al. 2005). 

It seems that both the features and overall shape of spectrum tend to change with Lutetia's rotation. Previously variations of spectral slope over the surface were found by Nedelcu et al.(2007) in the near-infrared wavelength range and by Busarev (2008) in the visible range. These data are related to the equatorial aspect and were interpreted as indicative of various surface mineralogy (Nedelcu et al. 2007). Our data corresponding to the aspect angle of 14 deg assume rather large surface heterogeneity of Lutetia to be seen in the integral observations from the near pole-on direction.

\section{Discussion}

Described results of photometric, polarimetric and spectral observations of Lutetia show that the use of different techniques produces a rather consistent picture of the main physical and optical properties of the asteroid: it appears to have a highly heterogeneous surface. The conclusion follows from 1) the non-zero lightcurve amplitude measured at the polar aspect; 2) spectral slope variations found both at the polar and equatorial aspects; 3) observed variations of polarization degree over the surface. These features could be explained by an assumption of a global non-convex shape (e.g. due to the presence of a large crater or craters) and heterogeneous surface texture and/or mineralogy. The hypothesis of the presence of a large crater in the northern hemisphere was also proposed by Carvano et al. (2008) to explain the value of Lutetia's albedo p$_V$=0.13 derived from their thermophysical model, that was smaller than the previous value of radiometric albedo p$_V$=0.22 obtained by Mueller et al.(2006). 

We have no strong evidence in favor of large albedo variegations over Lutetia's surface. Available radiometric measurements made at different aspects gave rather consistent values of Lutetia's albedo in the range of 0.19-0.22 with an estimated uncertainty of 0.02 (Tedesco et al. 2002, Mueller et al. 2006, Lamy et al. 2008). Our new estimation of the polarization albedo of 0.16$\pm$0.02 is still lower than radiometric albedo. However it was shown in section 2.2 that 
determination of Lutetia's albedo from polarimetric data has some difficulties due to particular polarization properties of this asteroid. The measured values of opposition effect and phase slope are consistent with moderate-albedo surface.  

Using our precise determination of absolute magnitude of Lutetia H=7.25 mag for the near polar aspect (corresponding to observations in 2004 and 2008) we calculated its albedo from available size estimations for these apparitions. The albedo ranges from 0.18 for the effective diameter of 110 km (Drummond et al. 2009) to 0.22 assuming the effective diameter of 100 km (Mueller et al. 2006). The above values of albedo correspond to zero phase angle and can not be directly compared to albedos of meteorites, usually measured at $\alpha$$\approx$3-5$^o$. We calculated so-called four-degree albedo as proposed by Shevchenko \& Tedesco (2006) using V(1,$\alpha$=4$^o$)=7.63$^m$. This value roughly corresponds to the absolute magnitude of the asteroid without taking into account the opposition surge. The four-degree albedo of Lutetia is in the range of 0.13-0.16 for the effective diameter in the range of 100-110 km. These values of albedo are consistent with particular types of carbonaceous chondrites and enstatite chondrites and are smaller than typical values for iron meteorites (e.g. Gaffey 1976). 
 
To compare albedo and spectral properties of Lutetia with laboratory measurements we need to take into account that they are greatly affected by particle size. On the basis of available data we expect that Lutetia's surface are covered by fine-grained regolith. The conclusion follows from 1) particular polarimetric properties of Lutetia characterizing by large inversion angle, and 2) behaviour of the emissivity spectra of Lutetia with a narrow 10 $\mu$m emission feature (Feierberg et al. 1983, Barucci et al. 2008). According to estimations at least a portion of Lutetia's surface should be covered by fine regolith with a grain size $\leq$20 $\mu$m. 

Fine-grained mixtures of components with different optical properties (irons, silicates, carbon) can drastically change spectral reflectivity suppressing silicate bands (e.g. Feierberg et al. 1982). The particle size is not well-controlled in laboratory measurements of crushed meteorites because of different fragility of their components. Moreover the processes that can affect the optical properties of regolith exposed to space are not enough understandable to confidently interpret asteroid spectra (see Chapman, 2004 for review). It is possible that the observed variations in spectral properties of Lutetia are related to different exposure history of its regolith due to large impact.

Both spectral and polarimetric observations show that Lutetia surface properties are quite different from those of most asteroids studied so far. In a new asteroid taxonomy Lutetia was classified in the Xc sub-class (DeMeo et al 2009),  very few members belong to this class, among them 97 Klotho presents spectral properties similar to those of Lutetia (Vernazza et al. 2009). Moreover, the polarization properties of Klotho (Belskaya et al. 2009) also resemble those of Lutetia and distinctly deviate from other moderate-albedo asteroids. We expect that these two bodies have a very similar surface composition. According to Vernazza et al. (2009), they are the best candidates to be the parent bodies of enstatite chondrites. This conclusion has difficulties to explain 1) the observed features in the Lutetia's visible spectra interpreted as indicative of aqueous alteration material (Lazzarin et al., 2004, 2009; Busarev 2004); 2) a presence of a 3 $\mu$m feature associated with hydrated minerals (Rivkin et al. 2000); 3) the features of 5.2-38 $\mu$m emissivity spectrum (Barucci et al. 2008); 4) particular polarization properties of Lutetia. The above mentioned features can be more naturally explained assuming similarity of Lutetia's surface to particular types of carbonaceous chondrites. In turn, this assumption requires an explanation of relatively flat spectral slope of Lutetia toward ultraviolet wavelength. Lazzarin et al. (2009) suggested several possible explanations but not excluded that available meteorite assemblages might not be representative of the Lutetia surface composition.  

All of the above mentioned data are related to the surface properties of Lutetia. To constrain the interior composition we need to estimate the mass and density of the asteroid. Although Prokof'eva-Mikhailovskaya et al. (2007) made a conclusion of a complex satellite system of Lutetia, no satellites are yet detected around the asteroid (Busch et al. 2009). The only available mass estimations of Lutetia come from the astrometric method and give a density comparable to iron meteorites (Baer et al. 2009). However available radar observations raise doubts as to the reliability of the estimated mass. The radar albedo of Lutetia span from 0.17$\pm$0.07 (Magri et al. 1999) to 0.24$\pm$0.07 (Shepard et al. 2008) and both exclude a metallic surface composition. Radar data are consistent with the composition similar to either enstantite chondrites or particular metal-rich CH type of carbonaceous chondrites (Shepard et al. 2010). A possible similarity with CO/CV composition is also not excluded within the available uncertainties.  

Observed variations of spectral and polarimetric properties over Lutetia's surface can be attributed not only to heterogeneity in surface texture but also in surface composition, e.g. due to contamination in a large impact. It might explain particular properties of Lutetia. However neither satellites nor family members have been yet found for this asteroid. Previously classified as a member of Nysa family (Williams 1989) Lutetia does not belong to any family in later classifications (e.g., Zappala et al. 1995). Further study of these questions is needed. Note that an existence of satellites smaller than 6 km in diameter is not excluded by available observations (Busch et al. 2009). 

\section{Conclusions}

On the basis of a detailed analysis of new photometric, polarimetric and spectral data on the asteroid 21 Lutetia,  together with observational data available in literature, we can draw some conclusions which can be checked during Rosetta fly-by:	

1. Lutetia has a non-convex shape, probably due to the presence of a large crater, and heterogeneous surface properties probably due to variations of texture and/or mineralogy related to surface morphology. 

2.  At least part of Lutetia's surface is covered by regolith composed of particles having a mean grain size less than 20 $\mu$m. 

3. The closest meteorite analogues of Lutetia's surface composition are particular types of carbonaceous chondrites {CO, CV, CH). It is also possible that Lutetia has specific surface composition not representative among studied meteorites or has a mixed mineralogy, e.g. due to surface contamination.     

Flyby observations of Lutetia by the Rosetta spacecraft in July 2010 will provide ground truth for Earth-based remote sensing. 

\begin{acknowledgements}
The research of INB has been supported by a Marie Curie Fellowship of the European Community. 
RG-H acknowledges the support from CONICET through grant PIP114-200801-00205, and the
partial financial support by CICITCA, Universidad Nacional de San Juan, through a research grant. We thanks to B. Carry for providing results on Lutetia's rotation properties prior to publication.
\end{acknowledgements}

\end{document}